\documentclass[conference]{IEEEtran}
\IEEEoverridecommandlockouts
\usepackage{cite}
\usepackage{amsmath,amssymb,amsfonts}
\usepackage{graphicx}
\usepackage{epstopdf}
\usepackage{subfigure}
\usepackage{stfloats}
\usepackage{textcomp}
\usepackage{xcolor}
\usepackage{flushend}
\usepackage{array}
\usepackage{algorithm}
\usepackage{algpseudocode}
\usepackage{multirow}
\def\BibTeX{{\rm B\kern-.05em{\sc i\kern-.025em b}\kern-.08em
    T\kern-.1667em\lower.7ex\hbox{E}\kern-.125emX}}

\DeclareMathOperator{\sinc}{sinc}

\begin{document}

\title{Enhance Security of Time-Modulated Array-Enabled Directional Modulation by Introducing Symbol Ambiguity}

\author{\IEEEauthorblockN{Zhihao Tao, Zhaoyi Xu, and Athina Petropulu}
\IEEEauthorblockA{
Department of Electrical and Computer Engineering, Rutgers University, Piscataway, NJ 08854\\
Email: \{zhihao.tao, zhaoyi.xu, athinap\}@rutgers.edu}
\thanks{This work was supported by ARO grants W911NF2320103.}
}

\maketitle

\begin{abstract}
In this paper, if the time-modulated array (TMA)-enabled directional modulation (DM) communication system can be cracked is investigated and the answer is YES! We first demonstrate that the scrambling data received at the eavesdropper can be defied by using grid search to successfully find the only and actual mixing matrix generated by TMA. Then, we propose introducing symbol ambiguity to TMA to defend the defying of grid search, and design two principles for the TMA mixing matrix, i.e., rank deficiency and non-uniqueness of the ON-OFF switching pattern, that can be used to construct the symbol ambiguity. Also, we present a feasible mechanism to implement these two principles. Our proposed principles and mechanism not only shed light on how to design a more secure TMA DM system theoretically in the future, but also have been validated to be effective by bit error rate measurements.
\end{abstract}


\section{Introduction}
Physical layer (PHY) security, which can date back to Shannon's secrecy analysis \cite{Shannon1949Comm} and Wyner's wiretap channel \cite{Wyner1975Wire}, has always been of a great interest to both academia and industry, especially with the developing of integrated sensing and communication systems that exposes communication information more readily to illegal users \cite{xu2023bandwidth, zhaoyi2022TMA}. There exist many approaches to achieve PHY security, one of which is to use directional modulation (DM). As the name suggests, DM transmits digitally modulated information signals only along the pre-selected spatial directions where the legitimate users are located, while distorts waveforms along all other unwanted directions \cite{daly2009dire}. As a very promising technique, DM does not require eavesdropper channel state information and cryptographic keys or impose interference on communication users, and hence has been brought into focus in enhancing PHY security over the past decade \cite{Xiao2023Synthesis}.

Generally, DM can be implemented in a synthesis way or synthesis-free way. The synthesis of DM transmitter entails the calculation of array excitation vectors and can be constructed via a orthogonal vector approach, symbol-level precoding, reconfigurable phase shifters, and so on\cite{ding2014vector, Ottersten2016, daly2009dire}. In contrast, the synthesis-free DM does not require stringent designs of amplitude and phase excitations and can be made possible by carefully constructing transmitter hardware, like the proposed retrodirective DM array, circular DM array, and antenna subset transmission-enabled array \cite{ding2017free, ding2017Circular, Hamdi2016subset}.

By introducing an additional degree of freedom, i.e., time, for array designs, time-modulated array (TMA) can also be used to realize a DM transmitter. In \cite{nie20144d} and \cite{Massa20144d}, the authors present a novel TMA-based four-dimensional antenna array to achieve DM functionality. In \cite{Massa2018time}, A synthesis-based TMA DM is proposed, which is optimized by a binary genetic algorithm. Until recently, a time-modulated orthogonal frequency division multiplexing (OFDM) DM transmitter is developed in \cite{tvt2019time}, where the authors utilize the spatial frequency expansion characteristic of TMA to enable PHY security. Specifically, \cite{tvt2019time} adopts a periodic ON-OFF switching pattern for each array element to generate harmonic signals at subcarriers. By conceiving the time ON-OFF pattern properly, the information signals received by legitimate uses at specific directions can be demodulated accurately by fast Fourier transform (FFT), while the OFDM symbols received at undesired directions will be scrambled since the symbols of each subcarrier are mixed with the harmonic signal from all other subcarriers. This scrambling scheme can be represented by a mixing matrix, which is of Toeplitz structure. As compared to other DM arrays, the proposed TMA OFDM DM transmitter in \cite{tvt2019time} exhibits attractive benefits, such as requiring only one radio frequency (RF) chain, FFT and inverse FFT (IFFT) compatible, synthesis free, etc., thus making it gain a lot of attention in recent years.

However, previous studies on the TMA DM technique mainly focus on its hardware implementation, energy efficiency improvement, time ON-OFF pattern designs and its applications \cite{tvt2019time, Wu2022Metamaterial, Xiao2023Synthesis, Massa2018time, li2022chaotic, shan2021multi, shan2022target}, and ignore if the TMA DM system is security enough in essence. As \cite{li2022chaotic} alleges, the widely used TMA-based DM has weak security due to the limited randomness of periodic time modulation pattern. In this paper, this point is validated for the first time. We first design a typical TMA OFDM DM transmitter as that of \cite{tvt2019time}, and then use a grid search algorithm to estimate the parameters of mixing matrix generated by this TMA transmitter. We discover empirically that the grid search can always find the actual parameters of the mixing matrix based on only one received scrambling symbol at the eavesdropper direction. Hence, the scrambling data transmitted along undesired directions can be defied by the estimated mixing matrix and the bit error rate (BER) will be as low as the legitimate directions. Following this observation, we propose introducing symbol ambiguity to the TMA OFDM DM system. Symbol ambiguity refers to that the eavesdropper cannot distinguish the actual transmitted OFDM symbols even though it obtains the estimated mixing matrix by grid search, or rather, there are more than one type of OFDM symbols that can satisfy the received scrambling data, so the eavesdropper cannot identify which one is the actual one. Next, we propose two principles for designing the TMA mixing matrix that can be used to construct the symbol ambiguity, i.e., rank deficiency and non-uniqueness of the ON-OFF switching pattern, and conceive a simple mechanism that is to rotate the TMA at a certain angle to fulfill these two principles. In this way, the defying of eavesdropper by grid search is defended and the security of TMA DM is enhanced. To some extent, our proposed principles and mechanism reveal how to design the TMA DM system theoretically in the future to enhance its PHY security in essence. Moreover, BER-based numerical results demonstrate the effectiveness of the proposed ideas, and the complexity of the defying of grid search is also analyzed.

The remainder of this paper is organized as follows. In Section II, we describe the system model of TMA OFDM DM transmitter. In Section III, we present how to crack the TMA DM system by grid search and how to defend this crack by our proposed schemes. Numerical results and analyses are provided in Section IV, and the followed is Section V, in which conclusions are drawn.

\section{System model of TMA-enabled DM}
In this section, we consider a TMA-enabled OFDM DM transmitter as proposed in \cite{tvt2019time}, where the adopted uniform linear array has $N$ elements spaced by half wavelength and the OFDM system comprises $K$ subcarriers spaced by $f_s$. The antenna array is connected to one RF chain and the input signals are OFDM symbols modulated by IFFT. Here we do not consider power allocation at the transmitter end or noise at the receiver end, so the power of each antenna in each subcarrier is set to be identical. Denote the digitally modulated sysmbol as $S_k$, where the subscript $k$ is the index of subcarrier. So a modulated OFDM symbol is give by
\begin{equation}
\boldsymbol{X}(t) = \frac{1}{\sqrt{K}} \sum \limits_{k=1}^K S_k \cdot e^{j2 \pi [f_0+(k-1)f_s]t},
\end{equation}
where $f_0$ denotes the frequency of first OFDM subcarrier. Note that we use boldface letters to denote vectors or matrices in this paper and we also eliminate the index of transmitted OFDM symbol here as it is appropriate to consider only one OFDM symbol for following analyses. Moreover, $S_k$ is usually normalized to be unit power and $1/\sqrt{K}$ in (1) is the power normalization coefficient.

Before being radiated into the half space, meaning the direction $\theta \in [0, \pi]$, the OFDM symbol needs to be multiplied by a antenna weight $w_n$ for the $n$-th element and manipulated by a ON-OFF switching function $U(t)$. Assume the half wavelength spacing of array elements is associated with $f_0$, we can obtain the following baseband signal transmitted along $\theta$
\begin{equation}
\boldsymbol{Y}(t, \theta) = \frac{1}{\sqrt{N}} \sum \limits_{n=1}^N \boldsymbol{X}(t) \cdot w_n \cdot U_n(t) \cdot e^{j(n-1) \pi \cos \theta},
\end{equation}
where $w_n$ usually satisfies 
\begin{equation}
w_n = e^{-j(n-1) \pi \cos \theta _{\rm 0}}.
\end{equation}
$\theta _{\rm 0}$ is the desired direction of legitimate user. The ON-OFF switching function $U_n(t)$ is usually designed as a square waveform, controlling the connect/disconnect between the $n$-th array element and RF chain. For simplicity, we use directly the normalized switch `on' time instant and the `on' time period to represent $U_n(t)$, which are denoted as $\tau _n^o$ and $\Delta \tau _n$, respectively. Then, expanding $U_n(t)$ in the form of Fourier series, we get
\begin{equation}
U_n(t) = \sum \limits_{m=-\infty}^{\infty} a_{mn} \cdot e^{j2m \pi f_s t},
\end{equation}
where
\begin{equation}
    a_{mn} = \Delta \tau _n \sinc (m \pi \Delta \tau _n) \cdot e^{-jm \pi (2 \tau _n^o + \Delta \tau _n)}.
\end{equation}
Here $\sinc (\cdot)$ is a unnormalized sinc function. By combining the above equations, we can obtain the transmitted OFDM symbol as
\begin{equation}
    \boldsymbol{Y}(t, \theta) = \frac{1}{\sqrt{NK}} \cdot \sum \limits_{k=1}^K S_k \cdot e^{j2 \pi [f_0+(k-1)f_s]t} \cdot \sum \limits_{m=-\infty}^{\infty} V_m,
\end{equation}
where
\begin{equation}\label{eq7}
\begin{split}
    V_m = & e^{j2m \pi f_s t} \sum \limits_{n=1}^N \Bigl(\Delta \tau _n \sinc (m \pi \Delta \tau _n) \\
    & e^{-jm \pi (2 \tau _n^o + \Delta \tau _n)} e^{j(n-1) \pi (\cos \theta - \cos \theta _{\rm 0})} \Bigl).
\end{split}
\end{equation}

By the orthogonal characteristic of OFDM, the received data in the $i$-th subcarrier can be written as
\begin{equation}
    {Y}_i (t, \theta) = \frac{1}{\sqrt{NK}} \sum \limits_{k=1}^K S_k \cdot e^{j2 \pi [f_0+(k-1)f_s]t} \cdot V_{m=i-k}.
\end{equation}
Using the FFT demodulation, we can further simplify (8) as $Y_i (t, \theta) = 1/\sqrt{NK} \cdot \sum_{k=1}^K S_k \cdot V_{m=i-k}^{'}$, where $V_m^{'}$ does not contain the term $e^{j2m \pi f_s t}$. Finally, (6) can be expressed in a matrix form, i.e., $\boldsymbol{Y} = \mathcal{T} \cdot \boldsymbol{S}$, where the mixing matrix $\mathcal{T}$ is
\begin{equation}\label{eq9}
    \mathcal{T} = \frac{1}{\sqrt{NK}} \left[ \begin{array}{ccccc}
V_0^{'}& V_{-1}^{'}& \cdots& V_{-(K-2)}^{'}& V_{-(K-1)}^{'}\\
V_1^{'}& V_0^{'}& \cdots& V_{-(K-3)}^{'}& V_{-(K-2)}^{'}\\
\vdots& \vdots& \ddots& \vdots& \vdots \\
V_{K-2}^{'}& V_{K-3}^{'}& \cdots& V_0^{'}& V_{-1}^{'}\\
V_{K-1}^{'}& V_{K-2}^{'}& \cdots& V_1^{'}& V_0^{'}
\end{array} \right],
\end{equation}
and $\boldsymbol{S} = [S_1, S_2, \cdots, S_K]$. Obviously, $\mathcal{T}$ is of Toeplitz structure. Now we need to design the ON-OFF switching function to implement DM functionality. To this end, $\tau _n^o$ and $\Delta \tau _n$ are chosen properly to satisfy $V_{m \ne 0}(\tau _n^o,\Delta \tau _n,\theta = \theta _0) = 0$ and $V_{m = 0}(\tau _n^o,\Delta \tau _n,\theta = \theta _0) \ne 0$, and the choice results are: 1) $\Delta \tau _n \in [0, 1], \tau _n^o \in \{\frac{h-1}{N}\}_{h=1,2,...,N}$ (note that the subscript $n$ is not necessarily equal to $h$); 2) $\tau _p^o \ne \tau _q^o, \Delta \tau _p = \Delta \tau _q$ for $p \ne q$; and 3) $\sum_{n=1}^N \Delta \tau _n \ne 0$. After this, we can find the the received OFDM signal along $\theta _0$ as $\boldsymbol{Y}(t, \theta _0) = \Delta \tau _n \cdot \sqrt{N/K} \cdot \boldsymbol{S}(t)$, and this can be recovered readily. As for $\theta \ne \theta _0$, it can be seen from (8) that the received data ${Y}_i (t, \theta)$ is scrambled by the symbols modulated onto all other subcarriers since the mixing matrix $\mathcal{T}$ is not a diagonal matrix any more. Therefore, PHY security is achieved by the TMA-enabled OFDM DM system.

\section{Defy and defend the TMA-enabled OFDM DM communication system}

\subsection{Proposed Defying Strategy Based on Grid Search}
Even scrambling, the eavesdropper can still received the data $\{\boldsymbol{Y}(t, \theta _e)\}_{t=1,2,...T}$ containing actual transmitted OFDM symbols, where $\theta _e$ is the eavesdropper direction, so one of the most interesting issues to us is if we (assume we are the eavesdropper now) can crack the TMA-enabled OFDM DM system by analyzing and processing these received data. In other words, we need to investigate if we can find the transmitted source symbol vector $\boldsymbol{S}(t)$ based on $\boldsymbol{Y}(t, \theta _e)$.

It is natural to consider adopting blind source separation methods, like class independent component analysis (ICA) algorithms. ICA do not require any other prior knowledge but observed data, and utilize independence and non-Gaussianity to separate the mixing matrix and source data. However, ICA cannot handle with phase and permutation ambiguity or scaling issues \cite{oja2000ica}, which is insignificant in most applications but unfortunately not in our problem. Another idea is to obtain estimated $\mathcal{T}(t)$ and $\boldsymbol{S}(t)$, i.e., $\hat{\mathcal{T}}(t)$ and $\hat{\boldsymbol{S}}(t)$, to minimize the error between $\boldsymbol{Y}(t, \theta _e)$ and $\hat{\boldsymbol{Y}}(t, \theta _e)$, where $\hat{\boldsymbol{Y}}(t, \theta _e) = \hat{\mathcal{T}}(t) \cdot \hat{\boldsymbol{S}}(t)$, and in this sense $\hat{\boldsymbol{S}}(t)$ is $\boldsymbol{S}(t)$ we want to find. We can use heuristic approaches like evolutionary algorithms or gradient descent-based methods to solve this minimization. But different from general optimization problems, in our case, we need to find the actual and only $\boldsymbol{S}(t)$ instead of locally optimal solutions, meaning that we have to eliminate the ambiguity of solved solutions. Otherwise, the eavesdropper cannot make sure that it finds the true transmitted source symbol. Therefore, we turn to global optimization algorithms, one of which is grid search.

Now let us elaborate how grid search works in defying the scrambling data. Since $\boldsymbol{S}(t)$ is transmitted randomly and independently, there will be no temporal correlations among $\{\boldsymbol{Y}(t, \theta _e)\}_{t=1,2,...T}$ that we can utilize, and hence we can consider only one received symbols, like $\boldsymbol{Y}(1, \theta _e)$, for the following analysis. Unless otherwise specified, the time index $t=1$ and $\theta _e$ are dropped hereafter out of brevity. Assume the transmitter adopts $Q$-order modulation and the eavesdropper knows $\theta _0$ and $\theta _e$, which is reasonable since they can be obtained by direction of arrival algorithms or analogue retrodirective technologies. So the total number of all combinations of source symbol modulated onto $K$ subcarriers is $Q^K$, and search space of these combinations is defined as $\mathcal{G}_S$. From Eq. (\ref{eq9}), we can know that the remaining parameters we need to search to determine $\mathcal{T}$ are $\{\tau _n^o\}_{n=1,2,...,N}$, $\{\Delta \tau _n \}_{n=1,2,...,N}$ and $N$. Assume we search potential values of $\tau _n^o$ and $\Delta \tau _n$ according to the above-mentioned choice results for $\tau _n^o$ and $\Delta \tau _n$ in Section II, we have the search space of $\{\tau _n^o\}_{n=1,2,...,N}$ as $\mathcal{G}_{\tau}$ and its size is $N!$. Set the $\{\Delta \tau _n \}_{n=1,2,...,N}$ as identical as $\Delta \tau$, and the stepsize of searching $\Delta \tau$ on [0, 1] is $1/L$, then define its search space as $\mathcal{G}_{\Delta \tau}$ and we can get the space size as $L$. Define the search space of $N$ as $\mathcal{G}_{N}$ and search it from 2 to $M$. Then we execute the grid search algorithm over $\mathcal{G}_{N}$, $\mathcal{G}_{\Delta \tau}$, $\mathcal{G}_{\tau}$ and $\mathcal{G}_S$ to form the estimated mixing matrix $\hat{\mathcal{T}}$ and the estimated source symbol vector $\hat{\boldsymbol{S}}$, and if the estimation error between $\boldsymbol{Y}$ and $\hat{\boldsymbol{Y}}$ reaches the threshold value $\epsilon$, it can be viewed that we find the actual $\mathcal{T}$ and $\boldsymbol{S}$. This grid search-based defying strategy is specified in Algorithm 1.
\begin{algorithm}[t]
  \caption{Grid Search-Based Defying Algorithm}\label{alg:grid}
  \hspace*{\algorithmicindent} \textbf{Input:} $\boldsymbol{Y}$, $\mathcal{G}_{N}$, $\mathcal{G}_{\Delta \tau}$, $\mathcal{G}_{\tau}$, $\mathcal{G}_S$, and $\epsilon$ \\
  \hspace*{\algorithmicindent} \textbf{Output:} $\hat{\mathcal{T}}$ and $\hat{\boldsymbol{S}}$
  \begin{algorithmic}[1]
  \For{$N$ in $\mathcal{G}_{N}$}
        \For{$\{\tau _n^o\}_{n=1,2,...,N}$ in $\mathcal{G}_{\tau}$}
            \For{$\Delta \tau$ in $\mathcal{G}_{\Delta \tau}$}
                \State Compute $\hat{\mathcal{T}}$ according to Eq. (\ref{eq7}) and Eq. (\ref{eq9});
                \For{$\boldsymbol{S}$ in $\mathcal{G}_S$}
                    
                    \State Compute $\hat{\boldsymbol{Y}} = \hat{\mathcal{T}} \cdot {\boldsymbol{S}}$;
                    \If{${\|\boldsymbol{Y} - \hat{\boldsymbol{Y}}\|_2}/{\sqrt{K}} \le \epsilon$}
                        \State $\hat{\boldsymbol{S}} = \boldsymbol{S}$;
                        \State \textbf{Return} $\hat{\mathcal{T}}$ and $\hat{\boldsymbol{S}}$.
                    \EndIf
                \EndFor
            \EndFor
        \EndFor
    \EndFor
  \end{algorithmic}
\end{algorithm}
Notice that the algorithm complexity of grid search-based defying is $O(NLN!Q^K)$.

By experiments, we find that the proposed grid search algorithm can always obtain the actual $\boldsymbol{S}$ and no ambiguity exhibits. We select some results to illustrate this point, which are shown in Fig. 1. Fig. 1 depicts the simulated BERs of TMA without defying and TMA defied by grid search against different transmitting directions, respectively. Here we set $N=4$, $k=6$, the actual $\Delta \tau = 1/N$, actual $\tau _n^o = (n-1)/N$, and adopt QPSK modulation for OFDM. The desired direction $\theta _0 = 60^{\circ}$, $L=10^4$ and the error threshold value $\epsilon = 10^{-5}$. Then we select the eavesdropper direction $\theta _e \in [106.2^{\circ}, 113.6^{\circ}]$ and the selecting stepsize is $0.2^{\circ}$. For these eavesdropper directions, we use the above defying algorithm to see if the actual $\mathcal{T}$ and $\boldsymbol{S}$ can be acquired based on only one received scrambling symbol $\boldsymbol{Y}(1, \theta _e)$, and the BER at $\theta _e$ will be very low or zero if they can be found. From Fig. 1, we can see that the BERs of undefied TMA are similar to the results shown in \cite{tvt2019time}, indicating that the TMA-enabled OFDM DM system can achieve PHY security in the desired direction. Meanwhile, we observe that the BERs at the eavesdropper directions are the same as the BER at $\theta _0$, which signifies that the scrambling data have been defied successfully by grid search. In the next section, more results will be shown to demonstrate the defying of TMA by our proposed method.
\begin{figure}[t]
\centerline{\includegraphics[width=3.4in]{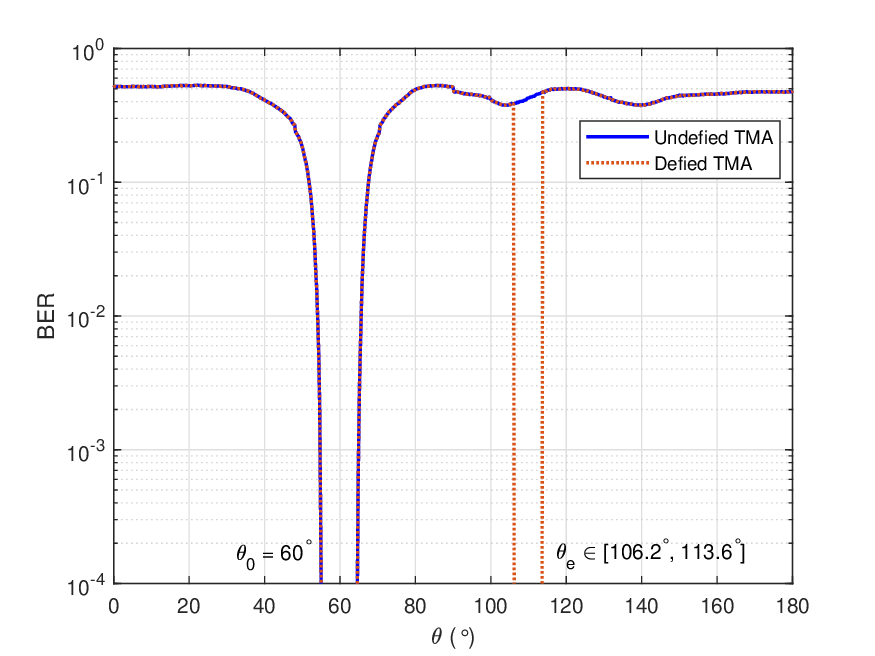}}
\caption{Simulated BERs of undefied and defied TMA versus $\theta$.}
\end{figure}

\subsection{Proposed Defending Principles and Mechanism}
Next we (assume we are the transmitter now) need to consider how to defend the above defying to enhance security of the TMA-enabled OFDM DM system. Since the proposed grid search algorithm can help the eavesdropper find the actual and only $\mathcal{T}$ and $\boldsymbol{S}$ and hence makes cracking the TMA DM system possible, we can introduce symbol ambiguity to the transmitter to impede it. That is to say, the eavesdropper will not identify the transmitted $\boldsymbol{S}$ when it is able to obtain multiple different $\hat{\mathcal{T}}$ and $\hat{\boldsymbol{S}}$ satisfying the error threshold value. The symbol ambiguity can keep the transmitter from being cracked in essence no matter how powerful the computational ability of eavesdropper is or how random the periodic time modulation pattern is.

There are two principles for designing the mixing matrix $\mathcal{T}$ that can be exploited to endow the TMA DM system with symbol ambiguity. One is rank deficiency. From $\boldsymbol{Y} = \mathcal{T} \cdot \boldsymbol{S}$, we can view the TMA OFDM DM system as a linear system. If $\mathcal{T}$ is rank deficient, i.e., non-full rank, this system will be underdetermined and have more than one solutions of $\boldsymbol{S}$ (it is impossible to have `no solution' since this is a physical system). The other one is non-uniqueness of the ON-OFF switching pattern, meaning there exist multiple ON-OFF switching patterns, i.e, multiple groups of $\Delta \tau $ and $\{\tau _n^o\}_{n=1,2,...,N}$, that can be found by grid search. In this way, our proposed defying method will obtain more than one $\hat{\boldsymbol{S}}$. In contrast, fining only one ON-OFF switching pattern is enough for the rank deficiency principle. To implement these two principles, we can design the transmitter from a theoretical point of view at first. For example, assuming $K=2$ and adopting BPSK modulation, we then formulate the following equations to fulfill a rank-deficient mixing matrix:
\begin{equation}
\begin{cases}
    & V_{m \ne 0}(N, w_n, \tau _n^o, \Delta \tau _n, \theta = \theta _0) = 0 \\
    & V_{m = 0}(N, w_n, \tau _n^o, \Delta \tau _n, \theta = \theta _0) \ne 0 \\
    & V_{m = 0}(N, w_n, \tau _n^o, \Delta \tau _n, \theta = \theta _e) = 0 \\
    & V_{m = -1}(N, w_n, \tau _n^o, \Delta \tau _n, \theta = \theta _e) = 0,
\end{cases}
\end{equation}
where we also take $N$ and antenna weights $w_n$ into account to design the transmitter achieving DM functionality meanwhile fulfilling ranking deficiency at the eavesdropper direction. Obviously, solving these equations is not easy or even impossible\footnote{We prefer to leave this to our future work.}.

\begin{figure}[t]
\centerline{\includegraphics[width=2.9in]{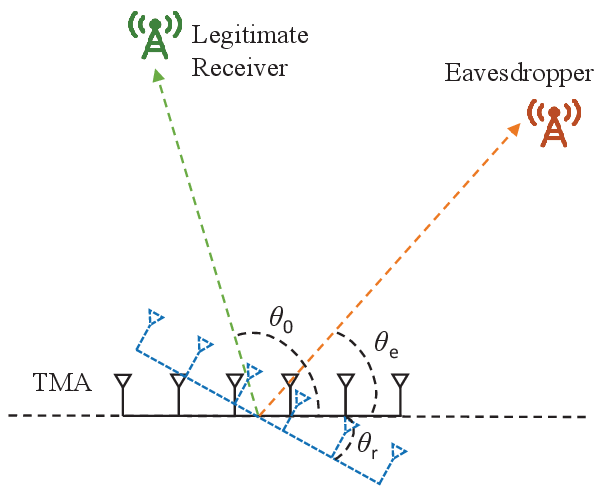}}
\caption{Illustration of the proposed defending mechanism.}
\end{figure}

Here we propose a simple mechanism that can realize these principles effectively, i.e., rotating the TMA DM transmitter at a certain angle $\theta _r$ that satisfies $\cos (\theta _e + \theta _r) - \cos (\theta _0 + \theta _r) = 2/N$, which is illustrated in Fig. 2. Take some examples to explain how it works. When $\cos (\theta _e + \theta _r) - \cos (\theta _0 + \theta _r) = 2/N$ is met, and let $\tau _n^o = (n-1)/N$, $N > K$, we can easily prove that $\mathcal{T}$ has the following form:
\begin{equation}
    \mathcal{T} = \frac{1}{\sqrt{NK}} \left[ \begin{array}{ccccc}
0& 0& \cdots& 0& 0\\
V_1^{'}& 0& \cdots& 0& 0\\
\vdots& \vdots& \ddots& \vdots& \vdots \\
0& 0& \cdots& 0& 0\\
0& 0& \cdots& V_1^{'}& 0
\end{array} \right],
\end{equation}
where $V_1^{'} \ne 0$. Clearly, $\mathcal{T}$ is rank deficient. Also, assuming $N=K=4$ and adopting BPSK modulation, we can obtain two different groups of $\Delta \tau $, $\{\tau _n^o\}_{n=1,2,...,N}$ and $\boldsymbol{S}$: $1/N$, $\{3/N, 1/N, 2/N, 0 \}$, $[+1, -1, -1, +1]$ and $3/N$, $\{0, 2/N, 3/N, 1/N \}$, $[-1, +1, +1, -1]$ that both correspond to the received $\boldsymbol{Y}$. Examining Eq. (8), we find that this non-uniqueness of ON-OFF switching pattern is exactly led by the characteristics of trigonometric functions, like periodicity and parity. In fact, the proposed mechanism is not limited to these listed examples and can be applied extensively. Further, $\cos (\theta _e + \theta _r) - \cos (\theta _0 + \theta _r) = 2/N$ can be extended to $\cos (\theta _e + \theta _r) - \cos (\theta _0 + \theta _r) = -2/N$ or others, like $\pm 4/N, \pm 6/N$, and so on; we mainly focus on the former for analysis in this paper. In a nutshell, $\cos (\theta _e + \theta _r) - \cos (\theta _0 + \theta _r) = 2/N$ provides an approach to introducing symbol ambiguity without painstakingly re-designing the time modulation pattern and antenna weights, and partial rationale behind it has been established mathematically.

\section{NUMERICAL RESULTS}
In this section, we present more numerical results to evaluate our proposed defying algorithm and defend mechanism.

First, we illustrate the algorithm complexity of proposed grid search-based defying strategy in Fig. 3 by showing the running time of finding the actual $\boldsymbol{S}$ with respect to different $N$ and $K$. For simplification, we put $N$ and $K$ together on the horizontal axis in Fig. 3, and their values are both from 2 to 9. We drop 1 for $N$ and $K$ since the TMA-enabled OFDM DM communication system cannot work for single antenna array or single carrier. The desired direction $\theta _0 = 60^{\circ}$ and the eavesdropper direction $\theta _e = 30^{\circ}$. $\Delta \tau$ is set as $1/N$, while $\{\tau _n^o\}_{n=1,2,...,N}$ is generated randomly according to $\tau _n^o \in \{\frac{h-1}{N}\}_{h=1,2,...,N}$. The values of $L$ and $\epsilon$ are selected properly here. Then, when we compare the running time of gird search against different $N$, $K$ is fixed as 2 and BPSK modulation is adopted, while $N$ is fixed as 3 and BPSK, QPSK are adopted, respectively, for the running time with different $K$. Note that these running time results are all based on one Intel Core i7-6700 CPU @ 3.4GHz and 16GB memory. From Fig. 3, we can observe that the running time required by grid search to find $\boldsymbol{S}$ grows fast with $N$ and $K$ increasing. When $N=9$, it needs about 20 hours to defy the scrambling data. Even though scaling up the number of antennas or subcarriers can slow the cracking, it cannot solve the security risks radically.

\linespread{1.3}
\begin{table*}[t]
\begin{center}
\caption{BER performance of the TMA DM system based on our proposed defying and defend methods}
\begin{tabular}{c|c|c|c|c|c|c|c|c|cc}
\hline
\hline
\multicolumn{1}{l|}{\multirow{2}{*}{Modulation Order}} & \multicolumn{1}{l|}{\multirow{2}{*}{$N$}} & \multicolumn{1}{l|}{\multirow{2}{*}{$K$}} & \multicolumn{1}{c|}{\multirow{2}{*}{$\Delta \tau$}} & \multicolumn{1}{c|}{\multirow{2}{*}{$\{\tau _n^o\}_{n=1,2,...,N}$}} & \multicolumn{1}{l|}{\multirow{2}{*}{$\theta _0$($^\circ$)}} & \multicolumn{1}{l|}{\multirow{2}{*}{$\theta _e$($^\circ$)}} & \multicolumn{1}{l|}{\multirow{2}{*}{Undefied BER}} & \multicolumn{1}{l|}{\multirow{2}{*}{Defied BER}} & \multicolumn{2}{c}{Defend}\\ \cline{10-11}
\multicolumn{1}{l|}{} & \multicolumn{1}{l|}{} & \multicolumn{1}{l|}{} & \multicolumn{1}{l|}{} & \multicolumn{1}{l|}{} & \multicolumn{1}{l|}{} & \multicolumn{1}{l|}{} & \multicolumn{1}{l|}{} & \multicolumn{1}{l|}{} & \multicolumn{1}{c|}{$\theta _r$($^\circ$)} & \multicolumn{1}{l}{Defended BER} \\
\hline
\multirow{3}{*}{BPSK} & 4 & 2 & 2/$N$ & {[}2/$N$, 3/$N$, 1/$N$, 0{]} & 80 & 40 & 0.2529 & 0 & \multicolumn{1}{c|}{-13.03} & 0.5000 \\
& 3 & 4 & 1/$N$ & {[}2/$N$, 1/$N$, 0{]} & 90 & 50 & 0.5031 & 0 & \multicolumn{1}{c|}{32.94} & 0.5000 \\
& 5 & 3 & 3/$N$ & {[}0, 2/$N$, 4/$N$, 1/$N$, 3/$N${]} & 60 & 30 & 0.8322 & 0 & \multicolumn{1}{c|}{5.60} & 0.4985 \\
\hline
\multirow{3}{*}{QPSK} & 4 & 2 & 2/$N$ & {[}2/$N$, 3/$N$, 1/$N$, 0{]} & 80 & 40 & 0.3124 & 0 & \multicolumn{1}{c|}{-13.03} & 0.5000 \\
& 3 & 4 & 1/$N$ & {[}2/$N$, 1/$N$, 0{]} & 90 & 50 & 0.4998 & 0 & \multicolumn{1}{c|}{32.94} & 0.5000 \\
& 5 & 3 & 3/$N$ & {[}0, 2/$N$, 4/$N$, 1/$N$, 3/$N${]} & 60 & 30 & 0.6995 & 0 & \multicolumn{1}{c|}{5.60}  & 0.5003 \\
\hline
\multirow{2}{*}{16QAM} & 3 & 2 & 1/$N$ & {[}2/$N$, 1/$N$, 0{]} & 90 & 50 & 0.5314 & 0 & \multicolumn{1}{c|}{7.06} & 0.3770 \\
& 4 & 2 & 2/$N$ & {[}2/$N$, 3/$N$, 1/$N$, 0{]} & 120 & 80 & 0.4190 & 0 & \multicolumn{1}{c|}{33.03} & 0.3745 \\
\hline
\end{tabular}
\end{center}
\end{table*}
\linespread{1.16}

\begin{figure}[t]
\centerline{\includegraphics[width=3.3in]{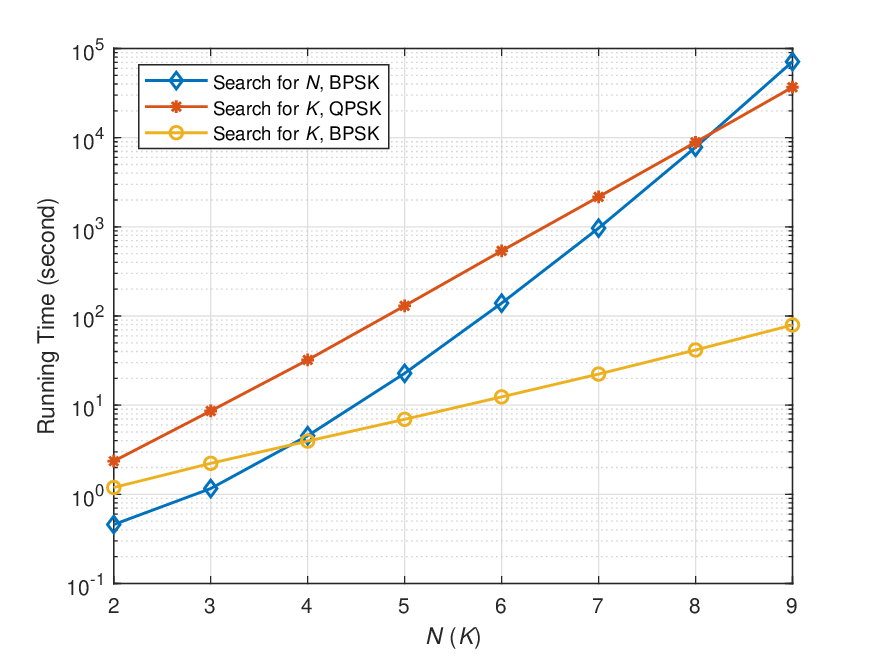}}
\caption{Running time of the proposed defying algorithm versus different $N$ and $K$.}
\end{figure}

Next, we exhibit the BER results of defying and defending TMA based on our proposed methods for various parameter configurations, which are shown in Table I. In Table I, `Undefied BER', `Defied BER' and `Defended BER' denote the BER of TMA transmitting at $\theta _e$ without being defied, with being defied by grid search and with being defended by the proposed mechanism, respectively. The defended BER is computed based on its expectation. We can see from this table that the defied BERs are all 0, indicating the grid search-based defying algorithm work very well in different scenarios. Also, by rotating the TMA transmitter at a certain angle $\theta _r$, the BERs are all improved, which demonstrates that our proposed mechanism is feasible and effective to defend the defying of grid search.

Finally, we showcase how our proposed rotation mechanism introduces symbol ambiguity in Table II in a more explicit way, where the BPSK modulation is adopted and the first transmitted source symbol vector $\boldsymbol{S}$ along $\theta _e = 40^{\circ}$ is $[-1, -1]$. The desired direction $\theta _0$ is set as $80^{\circ}$, and other actual parameters of TMA are shown in Table II. We select the first received scrambling data at $\theta _e$ for executing the grid search algorithm based on the defending mechanism, and the estimated four groups of results are summarized in Table II. It can be observed that the first and the second group of results have the same $\Delta \tau$ and $\{\tau _n^o\}_{n=1,2,...,N}$ as the actual ones but different $\boldsymbol{S}$, unveiling this is caused by the rank deficiency principle. In addition, the third and the fourth group of defied results have different $\Delta \tau$, $\{\tau _n^o\}_{n=1,2,...,N}$ and $\boldsymbol{S}$ from the actual ones, implying it is led by the non-uniqueness principle. We also can see that the defied BERs are not zero any more, which signifies that the PHY security of TMA DM system is enhanced by our proposed defending mechanism.


\linespread{1.3}
\begin{table}[t]
\begin{center}
\caption{Defied results based on our proposed defending mechanism}
\begin{tabular}{c|c|c|c|c|c|c}
\hline
\hline
\multicolumn{2}{c|}{}
& $N$ & $\Delta \tau$ & $\{\tau _n^o\}_{n=1,2,...,N}$ & $\boldsymbol{S}$ & BER \\
\hline
\multicolumn{2}{c|}{Actual}
& 3 & 1/$N$ & {[}2/$N$, 0, 1/$N${]} & {[}-1, -1{]} & 0.4988 \\
\hline
\multirow{4}{*}{Defied} & 1st & 3 & 1/$N$  & {[}2/$N$, 0, 1/$N${]}   & {[}-1, -1{]} & 0.2457 \\
& 2nd & 3 & 1/$N$ & {[}2/$N$, 0, 1/$N${]}  & {[}-1, +1{]}  & 0.2457 \\
& 3rd & 3 & 2/$N$ & {[}0, 1/$N$, 2/$N${]}  & {[}+1, -1{]}  & 0.7542 \\
& 4th & 3 & 2/$N$ & {[}0, 1/$N$, 2/$N${]}  & {[}+1, +1{]}  & 0.7542 \\
\hline
\end{tabular}
\end{center}
\end{table}
\linespread{1.16}

\section{CONCLUSION}
In this paper, we investigated if the data scrambling generated by TMA-enabled OFDM DM communication systems can be cracked. We first designed a grid search-based defying algorithm and found that the eavesdropper can always obtain the actual transmitted source symbols by this algorithm. Then we proposed two principles and one mechanism to endow the TMA DM system with symbol ambiguity to defend the defying of grid search. Numerical results showcase the effectiveness of proposed defying algorithm while demonstrate that the proposed defending methods are promising to enhance the PHY security of TMA DM. Partial theoretical analyses are also provided and in the future it is worthwhile exploring further theoretically sound algorithms for realizing more secure TMA-enabled DM systems.

\bibliography{May23_Ref}
\bibliographystyle{IEEEtran}

\end{document}